\documentclass{article}
\pdfoutput=1
\usepackage{authblk}
\usepackage{ragged2e}
\usepackage{graphicx}
\usepackage{caption}
\usepackage[margin=0.5in]{geometry}
\usepackage{gensymb}
\usepackage{amsmath,amsfonts,amssymb}
\usepackage{graphicx}
\usepackage{setspace}
\usepackage{booktabs}
\usepackage{siunitx}
\usepackage{multirow}
\usepackage{xcolor,colortbl}
\usepackage{cite}
\usepackage{placeins}
\usepackage[colorlinks=true, allcolors=blue]{hyperref}

\makeatletter
\def\@maketitle{%
  \newpage
  \null
  \let \footnote \thanks
    {\huge \@title \par}%
    \medskip%
    {\large
    \begin{tabular}[t]{l}%
      \RaggedRight
      \@author
    \end{tabular}
    \par}%
  \par}
\makeatother

\newcommand{\keywords}[1]{
\medskip
Keywords: \textit{#1}
}

\renewenvironment{abstract}{
\small
\medskip\medskip
}

\newcommand{\threesubsection}[1]{
\medskip
\textit{#1}: 
}

\title{Synchrotron radiation-based tomography of an entire mouse brain with sub-micron voxels: augmenting interactive brain atlases with terabyte data}

\author[a]{Mattia Humbel\/}
\author[a,b]{Christine Tanner\/\thanks{%
	Email Address: christine.tanner@unibas.ch}}
\author[c]{Marta Girona Alarc\'{o}n\/}
\author[a,b]{Georg Schulz\/}
\author[d]{Timm Weitkamp\/}
\author[d]{Mario Scheel\/}
\author[c]{Vartan Kurtcuoglu\/}
\author[a,e]{Bert M\"{u}ller\/}
\author[a]{Griffin Rodgers\/}

\affil[a]{%
Biomaterials Science Center,
Department of Biomedical Engineering,
University of Basel,
Hegenheimermattweg~167B/C,
4123~Allschwil,
Switzerland%
}

\affil[b]{%
Core Facility Micro- and Nanotomography,
Department of Biomedical Engineering,
University of Basel,
Hegenheimermattweg~167B/C,
4123~Allschwil,
Switzerland%
}

\affil[c]{%
The Interface Group,
Institute of Physiology,
University of Zurich,
Winterthurerstrasse~190,
8057~Zurich,
Switzerland%
}

\affil[d]{%
Synchrotron \textsc{Soleil},
L'Orme des Merisiers,
91190~Saint-Aubin,
France%
}

\affil[e]{%
Biomaterials Science Center,
Department of Clinical Research,
University Hospital Basel,
Spital\-stras\-se~8/12,
4031 Basel,
Switzerland%
}

\begin{document}
\RaggedRight

\maketitle


\begin{abstract}
\setlength{\parindent}{0 pt}
\setlength{\parskip}{0.5 em}
Synchrotron radiation-based X-ray microtomography is uniquely suited for post mortem three-dimensional visualization of organs such as the mouse brain. Tomographic imaging of the entire mouse brain with isotropic cellular resolution requires an extended field-of-view and produces datasets of multiple terabytes in size. These data must be processed and made accessible to domain experts who may have only limited image processing knowledge. We present extended-field X-ray microtomography with \SI{0.65}{\micro\meter} voxel size covering an entire mouse brain. The \num{4495} projections from $\num{8}\times\num{8}$ offset acquisitions were stitched to reconstruct a volume of $\num{15000}^3$ voxels. The microtomography volume was non-rigidly registered to the Allen Mouse Brain Common Coordinate Framework v3 based on a combination of image intensity and landmark pairs. The data were transformed block-wise and stored in a public repository with a hierarchical format for navigation and overlay with anatomical annotations in online viewers such as \texttt{Neuroglancer} or \texttt{siibra-explorer}. This study demonstrates X-ray imaging and data processing for a full mouse brain, augmenting current atlases by improving resolution in the third dimension by an order of magnitude. The \num{3.3}-teravoxel dataset is publicly available and easily accessible for domain experts via browser-based viewers.
\end{abstract}

\section{Introduction}
\label{sect:intro}
Understanding the relationship between structure and function in the brain is a unique challenge and
requires high-resolution brain imaging \cite{lichtman2011challenges}. This is complicated by the fact that the length scales of structures of interest span many orders of magnitude, even for smaller mammalian species. The mouse brain has a width of about \SI{1}{\centi\meter}, while individual cells are typically \SIrange{1}{10}{\micro\meter} in size, and the width of synaptic connections can be below \SI{100}{\nano\meter} \cite{lichtman2011challenges}.
Three-dimensional mapping of the brain's cytoarchitecture over these length scales requires complementary imaging modalities.
X-ray micro- and nanotomography are promising techniques for volumetric imaging of physically soft tissue, with hard X-ray microtomography offering micrometer or sub-micrometer resolution in a non-destructive fashion \cite{khimchenko2016extending,hieber2016nucleolar,dyer2017quantifying,topperwienThreedimensionalVirtualHistology2018b,topperwienCorrelativeXrayPhasecontrast2020b,chourrout_virtual_2023} and X-ray nanoholotomography allowing for imaging of structures down to tens of nanometers in size \cite{khimchenkoHardXRayNanoholotomography2018a,kuan2020denseneuronal}.
While these techniques have been successfully employed for brain tissue, usually only sub-volumes on the order of \SI{10}{\cubic\milli\meter} are accessible, compared to a volume of \SI{500}{\cubic\milli\meter} for the full mouse brain \cite{maThreedimensionalDigitalAtlas2005}.
Bridging this gap towards full-brain microtomography at sub-micrometer resolution has become an active area of research in recent years \cite{vescovi2018tomosaic,du2018extended,miettinen2019nrstitcher,vo2021dataprocessingmethods,foxleyMultimodalImagingSingle2021c}.

The field-of-view (FOV) offered by X-ray microtomography detector systems is limited by the number of pixels and the desired pixel size. Taking the Hamamatsu Orca Flash 4.0 V2 detector in use at the \textsc{Anatomix} beamline of Synchrotron \textsc{Soleil} \cite{weitkampTomographyBeamlineANATOMIX2017,weitkampMicrotomographyANATOMIXBeamline2022} as an example, the camera has $2048 \times 2048$ pixels and a pixel size of \SI{6.5}{\micro\meter}. With a \num{10}$\times$ magnifying objective, this gives a FOV of \SI{1.3}{\milli\meter}$\times$\SI{1.3}{\milli\meter}. To cover the entire mouse brain, the FOV needs to be extended.
Extending the FOV in the vertical direction, i.e., parallel to the tomographic rotation axis, is possible by translation of the specimen along the rotation axis, either in a helical scan, or by acquiring individual height steps and stitching the reconstructed volumes together. For lateral extension of the FOV, mainly two approaches have been pursued. One approach is to acquire individual tomograms and moving the sample on top of the rotation stage between scans, thus covering the entire cross section. The volumes are then stitched after reconstruction \cite{kyrieleisImageStitchingStrategies2009a,miettinen2019nrstitcher}.
An alternative scheme is to displace the rotation axis laterally with respect to the detector, and acquiring images over 360 degrees for each rotation axis position, which can then be stitched to produce extended-field projections \cite{vescoviRadiographyRegistrationMosaic2017,vescovi2018tomosaic}.
The first approach has the advantage that it allows for standard reconstruction of each tomogram.
However, non-rigid stitching of volumes needs to be established in case of sample deformation during the acquisition \cite{miettinen2019nrstitcher}. Artifacts resulting from the processing of truncated sinograms (also known as the \emph{local tomography} problem) need to be corrected for \cite{kyrieleisRegionofinterestTomographyUsing2011,silvaQuantitativeRegionofinterestTomography2018,robischIterativeMicrotomographyBiopsy2020}. The second approach
requires dedicated software for projection stitching and large-volume reconstruction \cite{vescoviRadiographyRegistrationMosaic2017,vescovi2018tomosaic,vo2021dataprocessingmethods}. The tilt angles of the rotation stage need to be aligned more precisely with respect to both the beam direction and the detector pixel rows than for standard microtomography scans.
The second approach has, however, been shown to be more dose- and time-efficient \cite{du2018extended}.

The resulting dataset covering the volume of the entire mouse brain with \SI{0.65}{\micro\meter} voxel size will require \SI{3}{\tera\byte} of storage space at 16-bit precision.
Sharing such a large image with the neuroscience community brings challenges in terms of data access, navigation, registration, and segmentation.
To leverage existing brain atlas annotations and thus allow the community to better navigate the large dataset, the microtomography volume must be registered to an atlas. 
Yet the registration of large images often proves challenging, as it can run into memory limitations or face excessive runtimes. 
To address these issues, we developed a distributed multi-resolution approach, where the large images are divided hierarchically into regions to circumvent memory limitations.
This approach was used to register a multi-terabyte sized microtomography volume to the Allen Mouse Brain Common Coordinate Framework v3 (CCFv3) \cite{wangAllenMouseBrain2020}. 

The effective sharing of research data within multidisciplinary communities requires easy access and interpretability. For imaging data, this ideally means online tools for visualizing, sharing, and annotating images instead of large repositories with only data download options.  
Recently, interactive visualization platforms have been developed for displaying volumes of mega- to petavoxel size, including \texttt{siibra-\allowbreak explorer}~\cite{siibra} and \texttt{Neuroglancer}~\cite{neuroglancer}.
\texttt{siibra-explorer} is a browser-based viewer for brain atlases, that allows seamless querying of semantically and spatially anchored datasets thanks to tight integration with the Human Brain Project Knowledge Graph. 
\texttt{Neuroglancer} is a multi-resolution viewer capable of displaying tera- to petavoxel large datasets and their segmentations fast enough to be practical. The location and orientation of any view are encoded in the uniform resource locator (URL), which can be shared for easy collaboration. As open-source software, \texttt{Neuroglancer} is highly extensible. Its usefulness for collaborative brain studies has been demonstrated, for example, for segmenting the \href{https://neuroglancer-demo.appspot.com/fafb.html#!gs://fafb-ffn1/main_ng.json}{full adult fly brain} dataset at $\num{4} \times \num{4} \times \SI{40}{\cubic\nano\meter}$ resolution, resulting in \SI{115}{\tera\byte} of image data~\cite{zhengCompleteElectronMicroscopy2018}.
Given these promising properties, we converted the registered microtomography volume of an entire mouse brain to the efficient pre-computed \texttt{Neuroglancer} format, which can also be read by \texttt{siibra-explorer}, and then uploaded it to \textsc{Ebrains}~\cite{ebrains} for dissemination.

This work builds on the reconstruction pipeline for extended field-of-view tomographic reconstruction of entire organs with cellular resolution initially presented by Rodgers and co-workers~\cite{rodgers2022mosaic}. Here, we report on (i) the acquisition of a large microtomography volume of an entire mouse brain with pixel size of \SI{0.65}{\micro\meter} and its (ii) reconstruction, (iii) registration to the Allen Mouse Brain CCFv3 atlas, (iv) conversion to hierarchical \texttt{Neuroglancer} format for fast interactive visualization and access, and (v) dissemination via the \textsc{Ebrains} Data and Knowledge services. This enables the imaging of an entire mouse brain with micrometer resolution in all three spatial dimensions and efficient sharing of the microtomography volume.

\section{Results and Discussion}
\label{sec:resultsdiscussion}

\subsection{The full mouse brain dataset}
\begin{figure}[htbp]
\centering
\includegraphics[width = 0.9\textwidth]{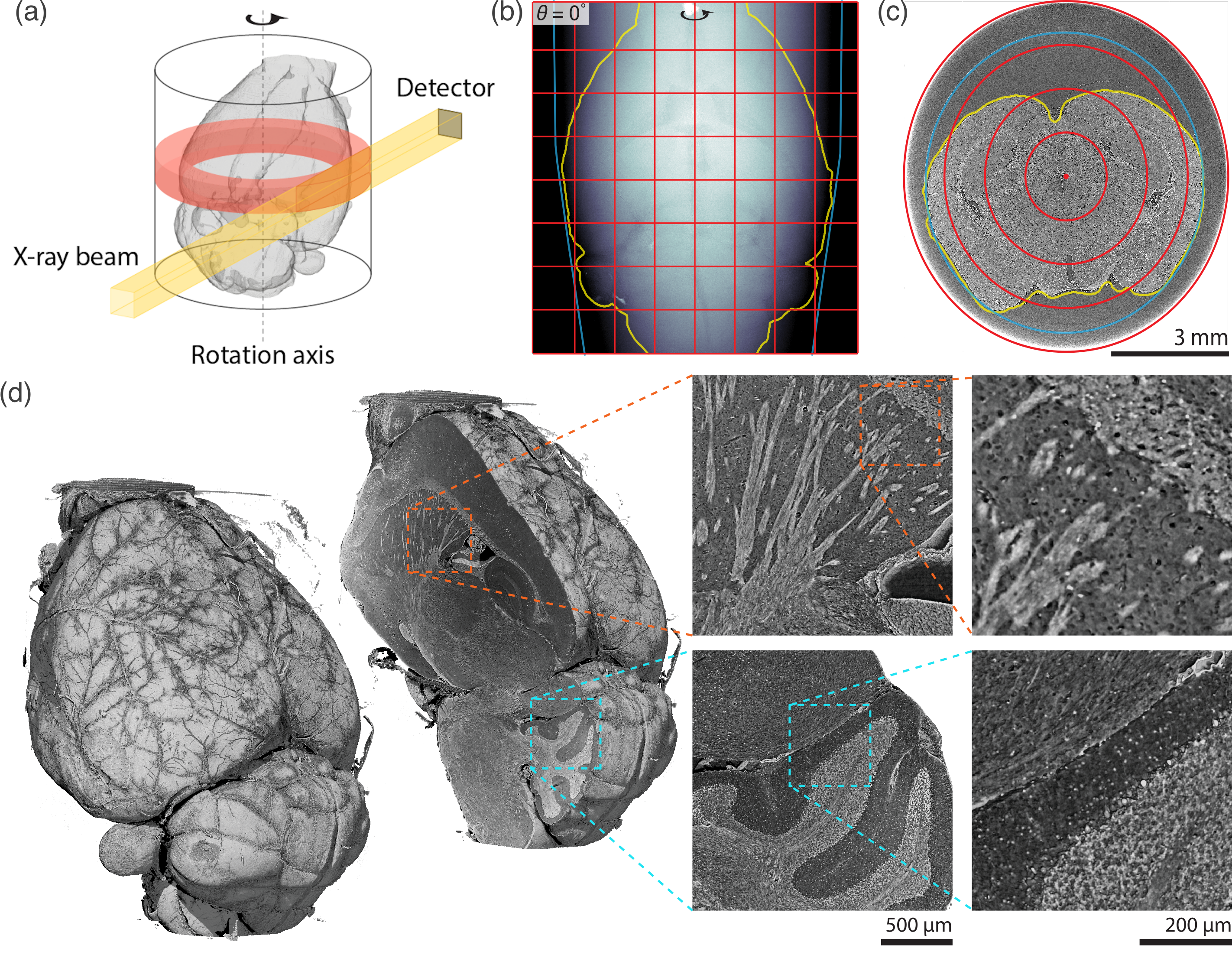}
\caption{%
Large field-of-view (FOV) X-ray microtomography of the entire mouse brain. To successively illuminate the entire cross-sectional area, the rotation axis was displaced relative to the detector and the X-ray beam (a).
An exemplary stitched projection is shown at \ang{0} rotation angle for the acquisition of the mouse brain with \num{8}$\times$\num{8}-times extended FOVs (b). Red lines delineate single \num{2048}$\times$\num{2048}-pixel detector FOVs that were combined with an overlap of \num{200} pixels to form a single \num{14982}$\times$\num{14982}-pixel projection. A reconstructed slice is shown with corresponding stitching positions in red (c). Yellow lines indicate the outline of the mouse brain and blue lines indicate the inner wall of the Eppendorf container used as a sample holder (b, c).
After reconstruction, the volumetric data can be viewed along any virtual slicing direction (d). Views from a sagittal slice are shown, alongside enlarged regions in the caudoputamen (orange) and cerebellum (cyan).
}
\label{fig:acquisition_rendering} 
\end{figure}

The full mouse brain, defined here
to include the cerebellum, but not the olfactory bulbs,
was scanned with eight height steps, see Figure~\ref{fig:acquisition_rendering}b (caudal part of olfactory bulbs at top, top of brain stem at the bottom). The resulting stitched dataset contained $\num{14982} \times \num{14982} \times \num{14784}$ voxels, or \num{3.3} teravoxels. At \num{16}-bit precision, this corresponds to a data size of \SI{6.6}{\tera\byte} (where $\SI{1}{\tera\byte} = 10^{12} \text{ bytes}$). It should be noted that this reflects the entire reconstructed FOV of \SI{911}{\cubic\milli\meter}. Re-orienting and cropping to more closely cover the approximately \SI{500}{\cubic\milli\meter} volume of the mouse brain would reduce data size.

Figure~\ref{fig:acquisition_rendering}d shows an overview rendering of the entire mouse brain data with $\num{8} \times \num{8} \times \num{8}$ downsampling. Virtual sagittal sectioning as in Figure~\ref{fig:acquisition_rendering}d (right) reveals macroscopic regions such as the isocortex, olfactory areas, hippocampal formation, cerebral nuclei, fiber tracts, and cerebellum.
The full-resolution data provide in-plane details similar to conventional histology, but with the benefit of three-dimensional isotropic resolution. This allows for exploration of the volumetric data in any virtual slicing plane.

\subsection{Registration to Allen Mouse Brain Common Coordinate Framework}
Table~\ref{tab:bestRegParameters}
lists the mean registration errors from two-fold cross-validation based on \num{70} manually selected landmark pairs.
Results are shown for the configuration achieving the lowest errors for affine and non-rigid registration as well as the best configuration when excluding image gradients ($w_\textrm{$|\nabla \mathbf{V}|$}=0$), segmentations ($w_\textrm{S}=0$), and/or landmarks ($w_\textrm{L}=0$).  
Lowest errors were achieved for affine and non-rigid registration when landmarks were incorporated into the cost function, i.e.~$w_\textrm{L}>0$. The weights providing the lowest mean error for the two-fold cross-validation, shown in the last two columns in Table~\ref{tab:bestRegParameters},
were employed for the final registration. Note that the \num{70} selected landmarks were used without exception  for the final registration.

\begin{table}[htbp]
    \centering
    \caption{Accuracy of image registration based on landmark error $f_\mathrm{D}$ from two-fold cross-validation are given for weight configurations of image gradient ($w_\textrm{$|\nabla \mathbf{V}|$}$), segmentation ($w_\textrm{S}$), and landmarks ($w_\textrm{L}$). For non-rigid registration, meta-parameters include bending energy penalty weight ($w_\textrm{BE}$) and grid spacing ($s$) in voxels. 
    The lowest mean landmark error of \SI{0.08}{\milli\meter} was achieved for non-rigid registration with the configuration shown in the last column (gray).
    The other columns list the best configurations when excluding image gradients ($w_\textrm{$|\nabla \mathbf{V}|$}=0$), segmentations ($w_\textrm{S}=0$), and/or landmarks ($w_\textrm{L}=0$). For comparison, the mean landmark error after manual pre-alignment was \SI{2.49}{\milli\meter}.}
    \label{tab:bestRegParameters}
    \vspace{10pt}
    \begin{tabular}{@{\hskip 0mm}c@{\hskip 1mm}|cccccc|ccccc >{\columncolor[gray]{0.85}}c}
    \toprule
         & \multicolumn{6}{c|}{Affine registration} & \multicolumn{6}{c}{Non-rigid registration}\\
    \hline
      \rule{0pt}{2.25ex} $w_\textrm{$|\nabla \mathbf{V}|$}$ & 0 & 10$^{-8}$ & 10$^{-8}$ & 0 & 10$^{-8}$ & 10$^{-8}$ & 0 & 10$^{-8}$ & 10$^{-8}$ & 0 & 10$^{-8}$ & 10$^{-8}$\\
      $w_\textrm{S}$  & 0 & 0 & 1 & 0 & 0 & 1 & 0 & 0 & 1 & 0 & 0 & 1\\
      $w_\textrm{L}$  & 0 & 0 & 0 & 1 & 1 & 1 & 0 & 0 & 0 & 0.1 & 0.1 & 0.1\\
      \hline
      \rule{0pt}{2.25ex} $w_\textrm{BE}$ & - & - & - & - & - & - & 100 & 1000 & 100 & 1000 & 1000 & 1000\\
      $s$            & - & - & - & - & - & - & 16 & 16 & 16 & 16 & 16 & 16\\
      \hline
      & & & & & & & & & & & & \\
      \multirow{-2}{*}{{\shortstack{$f_\mathrm{D}$ [mm]}}} & \multirow{-2}{*}{0.25} & \multirow{-2}{*}{0.16}  & \multirow{-2}{*}{0.16} & \multirow{-2}{*}{0.14} & \multirow{-2}{*}{0.14} & \multirow{-2}{*}{0.14} & \multirow{-2}{*}{0.40} & \multirow{-2}{*}{0.13} & \multirow{-2}{*}{0.13} & \multirow{-2}{*}{0.09} & \multirow{-2}{*}{0.08} & \multirow{-2}{*}{\bf 0.08}\\
    \bottomrule  
    \end{tabular}
\end{table}

\begin{figure}[htb]
\centering
\includegraphics[scale=0.9]{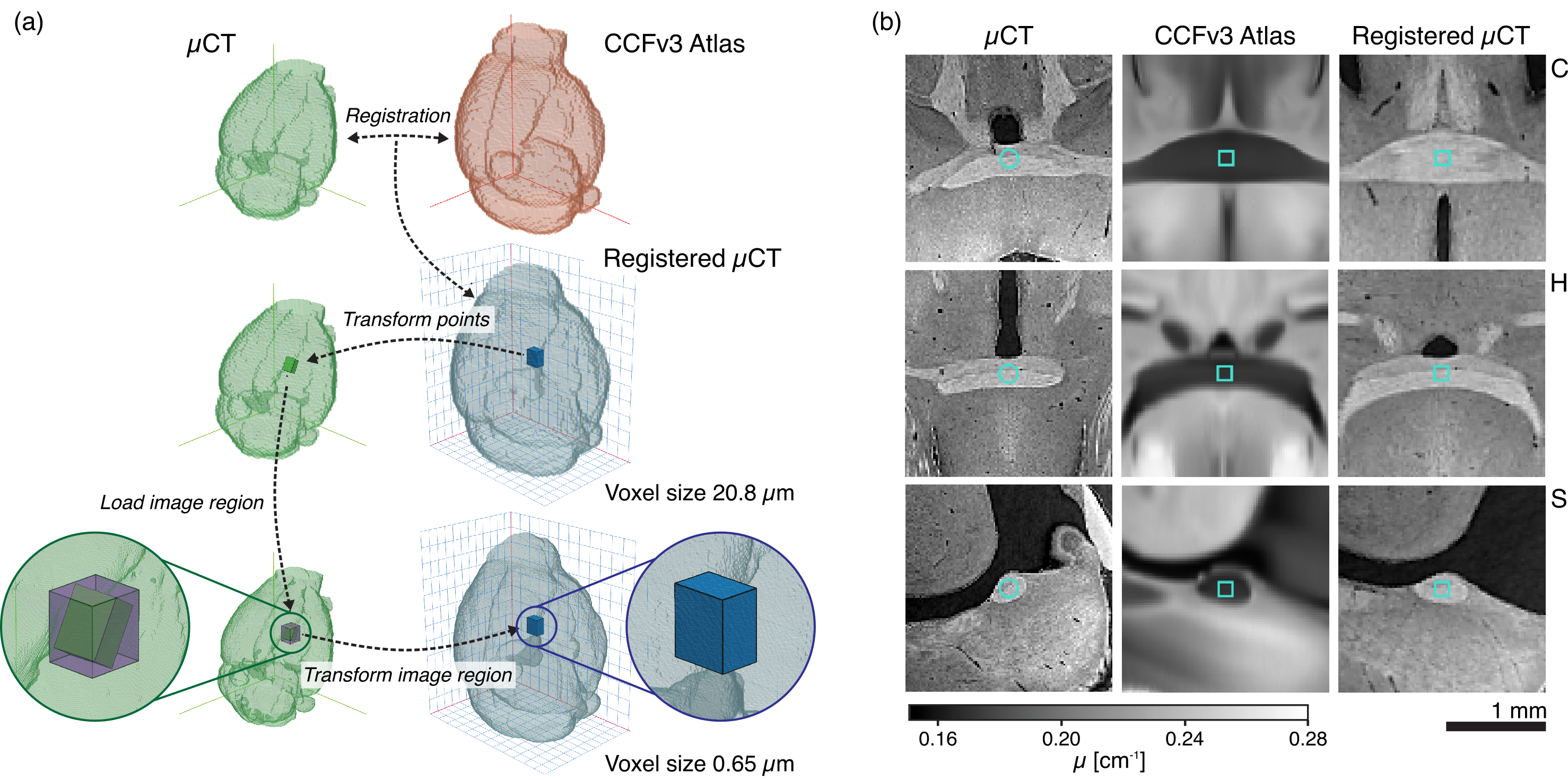}
\vspace{3mm}
\caption{%
(a) Illustration of the distributed hierarchical 3D image registration and transformation framework. (top) Registration was performed with the \num{32}$\times$ downsampled microtomography (\textmu CT) images to fit in available memory and have a resolution similar to that of the atlas. (middle) Given an axes-aligned local image region in atlas space (blue box), the resulting spatial transformation provided the corresponding coordinates in the \textmu CT image space (green box). (bottom) Local transformation of the original \textmu CT image was then performed by loading only the axis-aligned image region (purple box), which encloses the potentially skewed green box. Repeating this procedure for all $\num{12}\times\num{12}\times\num{12}$ local regions in the reference space created the registered \textmu CT image.
(b)
Registration results at corresponding landmark positions in the original microtomography, the atlas, and the non-rigidly registered \textmu CT datasets are displayed in virtual coronal (C), horizontal (H), and sagittal (S) planes. Representative landmark pairs were selected by Observer 1 (cyan). Image intensities for the \textmu CT and the two-photon microscopy-based template of the Allen Mouse Brain CCFv3 atlas were scaled to $[1,99]$ percentile of pixel intensities. The color bar indicating the linear attenuation coefficient $\mu$ applies to the \textmu CT images.
}
\label{fig:registration} 
\end{figure}

\begin{figure}[ht!]
\centering
\includegraphics[scale=0.78]{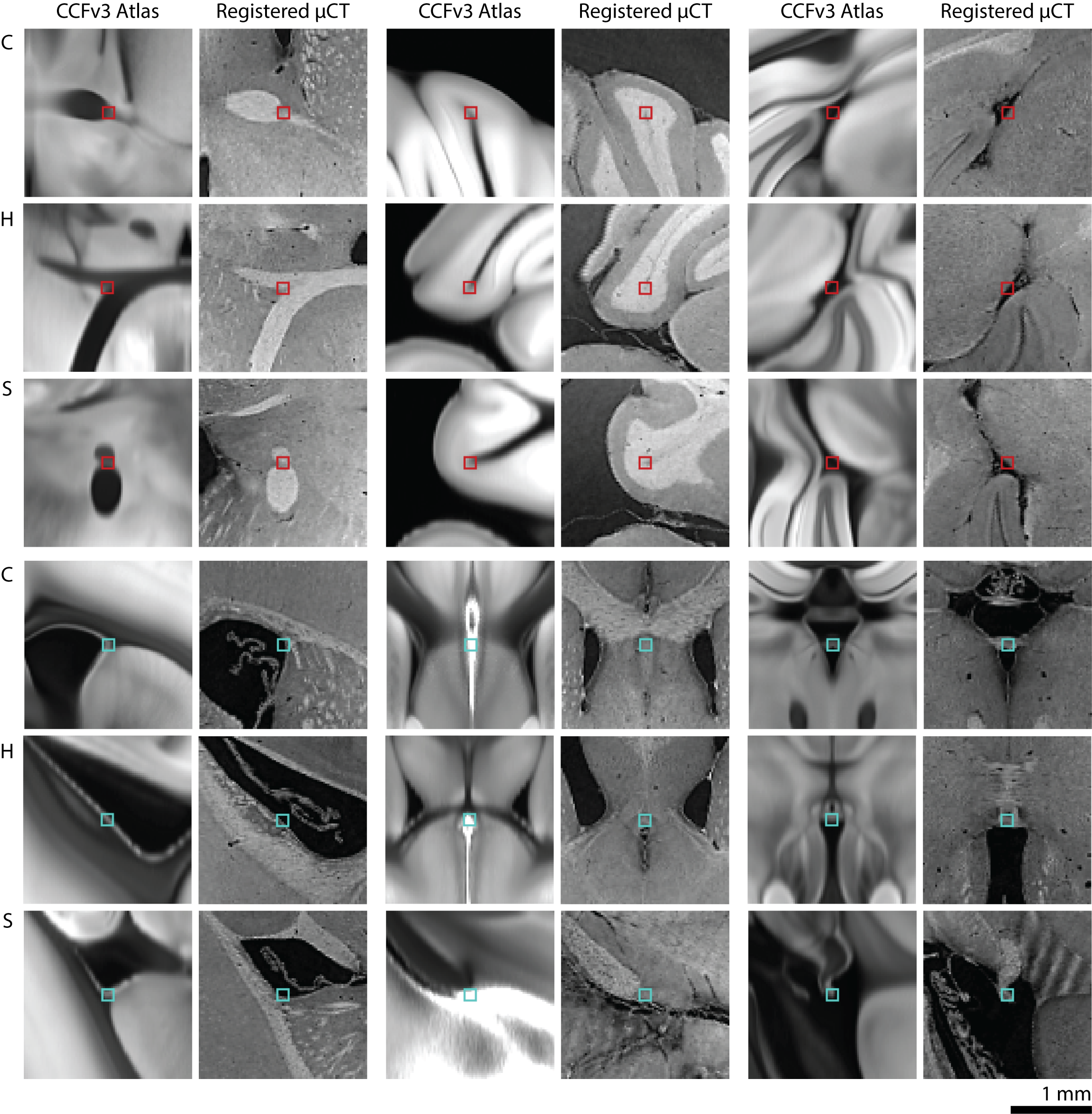}
\vspace{3mm}
\caption{
Registration results at corresponding landmark positions in the Allen Mouse Brain CCFv3 atlas and the non-rigidly registered microtomography (\textmu CT) datasets are displayed in virtual coronal (C), horizontal (H), and sagittal (S) planes. Representative landmark pairs selected by Observer 1 (bottom, cyan) and Observer 2 (top, red) are shown.
Image intensities for the \textmu CT and the two-photon microscopy-based template of the Allen Mouse Brain CCFv3 atlas were scaled to $[\num{1}, \num{99}]$ percentile of pixel intensities.}
\label{fig:regResultLMs2} 
\end{figure}

The number of confidently identifiable landmark pairs was limited by anatomical variations as well as differences in structure appearances between the microtomography volume and the average two-photon microscopy CCFv3 template.
Landmarks included ideal point landmarks as well as central positions of larger anatomical structures. 
Figure~\ref{fig:registration} (right) shows image regions around a manually selected landmark from Observer 1 in the original images and after registration. After non-rigid registration, reasonable alignment of corresponding anatomical structures can be seen. The combination of intensity-based registration and landmark alignment with optimized weighting supported robustness against small errors in manual landmark positions. Further landmarks and their alignment after registration can be seen in Figure~\ref{fig:regResultLMs2}.

\begin{figure}[htbp]
\centering
\includegraphics[width=0.78\textwidth]{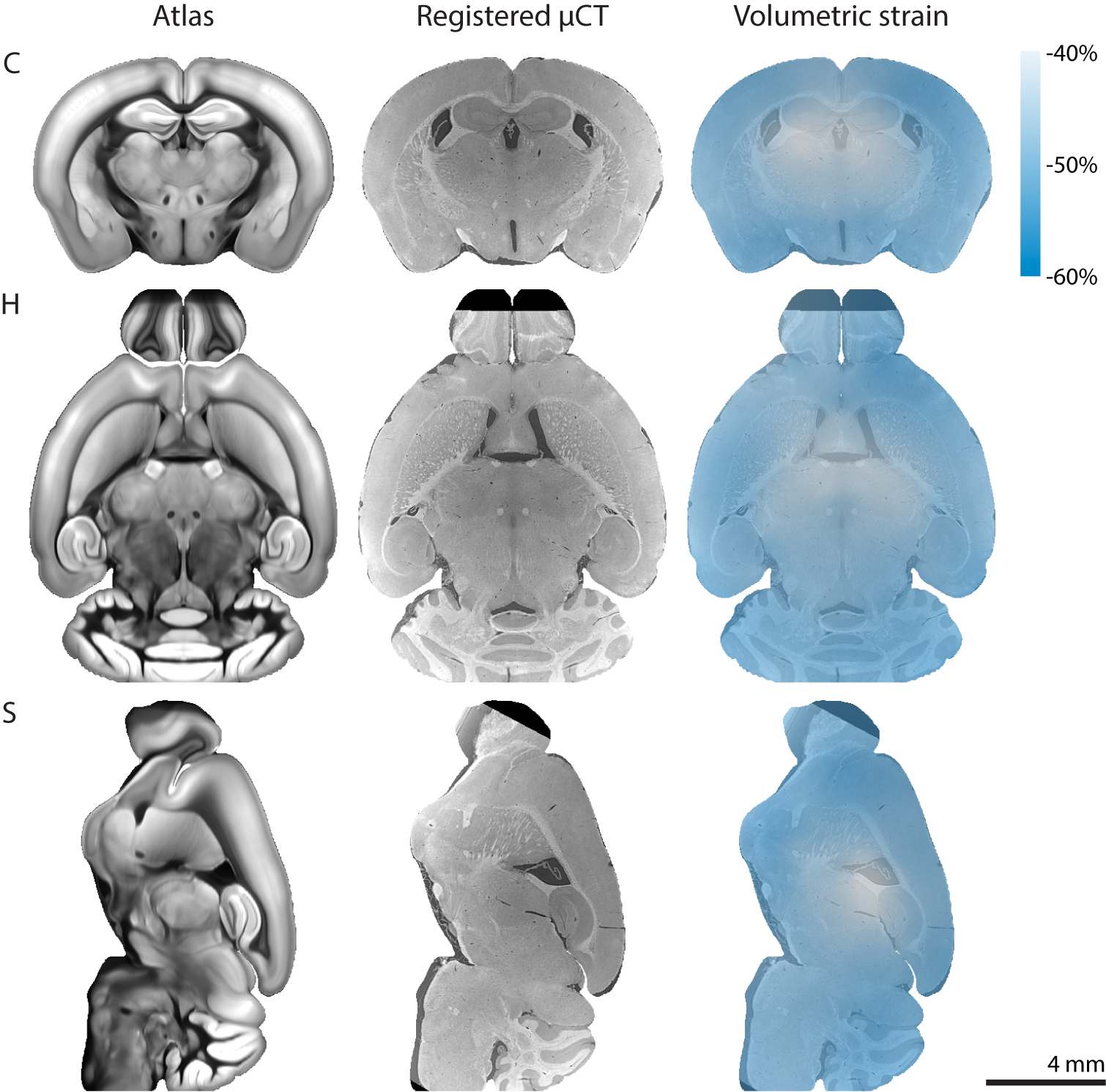}
\vspace{3mm}
\caption{Non-rigid registration results. Virtual coronal (C), horizontal (H), and sagittal (S) slices are shown from the two-photon microscopy-based template of the Allen Mouse Brain CCFv3 atlas~\cite{wangAllenMouseBrain2020}, the non-rigidly registered microtomography (\textmu CT) volume, and the volumetric strain as a result of registration.}
\label{fig:regResult1} 
\end{figure}

Figure~\ref{fig:regResult1} shows virtual slices from the atlas template image and the non-rigidly registered microtomography volume.
The resulting alignment enables the observation of similarities and differences in anatomy and appearance of structures.
Due to the ill-posed problem of registering multi-modal data across mice, the alignment is in some places imperfect. Yet, the alignment is sufficient to provide an understanding of the shown anatomical structures when overlaid with the atlas segmentation. Thus, the present registration is useful for general alignment with the Allen Mouse Brain CCFv3 atlas and navigation, but labels should not be transferred from the atlas without further refinement.

In the Allen Mouse Brain CCFv3 template, the whole brain volume is \SI{506}{\cubic\milli\meter}. This is because the underlying serial two-photon data of formalin-fixed agarose-immersed brains were scaled to match the unfixed fresh-frozen Nissl-based template of CCFv1~\cite{wangAllenMouseBrain2020}. This achieved backward compatibility and closeness in volume to in vivo mouse brains~\cite{ma2008vivo}. The mean brain volume in the original serial two-photon data used for the CCFv3 template was \SI{435}{\cubic\milli\meter}~\cite{wangAllenMouseBrain2020}.
By comparison, transforming the CCFv3 template to the coordinate frame of our mouse brain dataset in \SI{100}{\percent} ethanol produced a volume of \SI{244}{\cubic\milli\meter}.
The local change in volume as a result of non-rigid registration is shown in the right column of Figure~\ref{fig:regResult1}. A smooth shrinkage pattern with a mean volumetric strain of \SI{-48}{\percent} can be observed, with larger shrinkage on the outside of the brain.
We previously observed a median volumetric strain of \SI{-39}{\percent} from formalin to ethanol embedding~\cite{rodgersVirtualHistologyEntire2022a}, which here would correspond to an estimated volume of \SI{400}{\cubic\milli\meter} in formalin and thereby falls within the range of volumes reported for the \num{1675} mouse brains used for the CCFv3 template. 

\subsection{Exploring and disseminating the full mouse brain dataset}

\begin{figure}[htbp]
\centering
\includegraphics[width=0.75\textwidth]{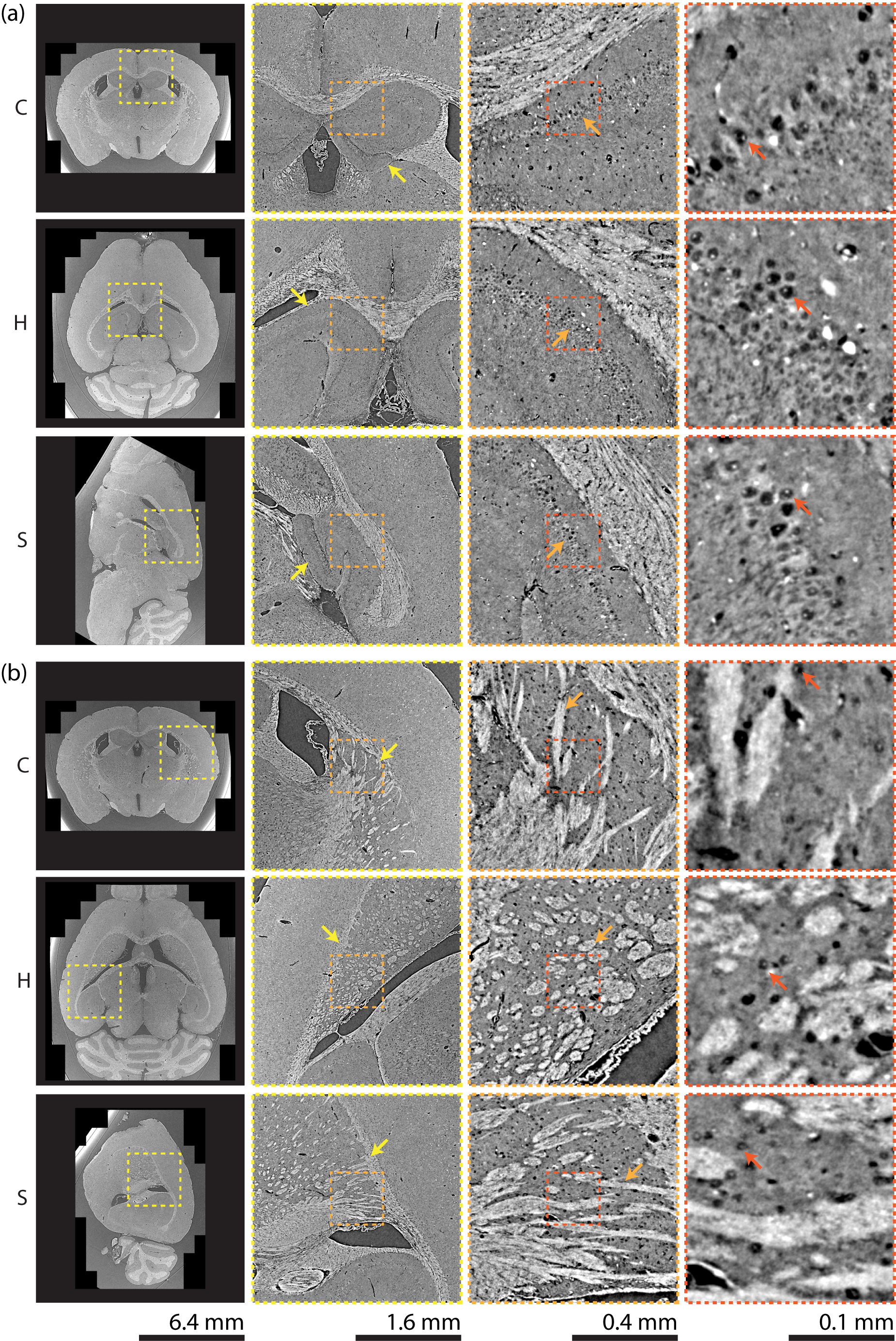}
\vspace{3mm}
\caption{Microanatomical details from the (a) hippocampal and (b) striatum dorsal regions are shown in orthogonal virtual coronal (C), horizontal (H), and sagittal (S) slices of the microtomography volume registered to the Allen Mouse Brain CCFv3 atlas. (a) Magnified views reveal structures of increasing detail, with arrows indicating (yellow) the hippocampus (including dentate gyrus and Ammon's horn), (orange) the CA1 field of Ammon's horn, and (red) the nucleus of an individual pyramidal neuron. (b) Details from the striatum dorsal region are shown with arrows indicating (yellow) caudoputamen, (orange) fiber tracts, and (red) the nucleus of a single neuron. Multi-resolution datasets are publicly shared in gzip-compressed pre-computed \texttt{Neuroglancer} format.}
\label{fig:hierarchicalZoomRoi1} 
\end{figure}

Details of microanatomical structures of the registered microtomography volume are shown in Figure~\ref{fig:hierarchicalZoomRoi1} for the hippocampal and striatum dorsal region.
At the highest resolution, cells, fibers, and other microscopic structures are clearly visible. The displayed virtual slices were downloaded from the publicly available dataset via \texttt{CloudVolume}, demonstrating that this dataset can not only be viewed, but also downloaded for processing.

\begin{figure}[htbp]
\centering
\begin{tabular}{cc}
\includegraphics[width=.476\textwidth]{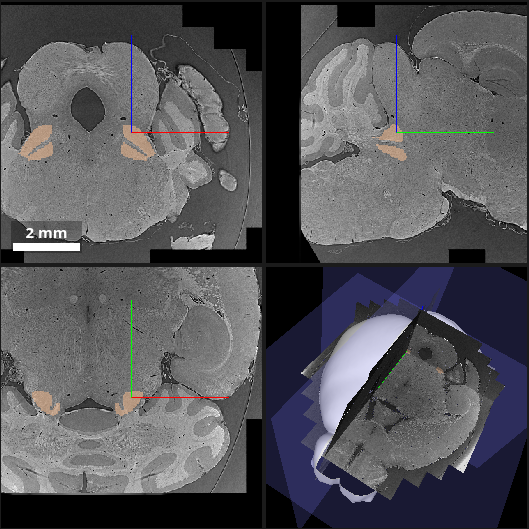} &
\includegraphics[width=.476\textwidth]{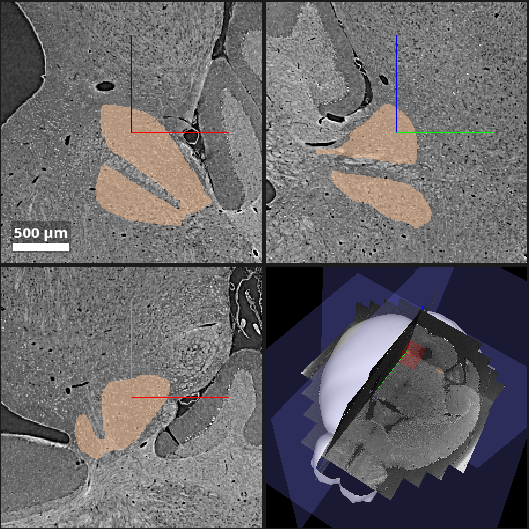}
\end{tabular}
\vspace{3mm}
\caption{Browser-based navigation of the publicly available mouse brain dataset. Visualization of microtomography image in the \texttt{siibra-explorer} is demonstrated with an overlaid segmentation of the parabrachial nucleus region from the Allen Mouse Brain CCFv3. The multi-resolution viewer allows for exploring the full dataset at lower resolutions (left), while still loading the full-resolution data at higher magnifications (right). Data are available at  \href{https://atlases.ebrains.eu/viewer/\#/a:juelich:iav:atlas:v1.0.0:2/t:minds:core:referencespace:v1.0.0:265d32a0-3d84-40a5-926f-bf89f68212b9/p:minds:core:parcellationatlas:v1.0.0:05655b58-3b6f-49db-b285-64b5a0276f83/@:0.0.0.-W000.._Ycgy.2-QZxH._cBFA.ys8lC..mr0..55K0.EXxe~.ZN7..1OS/vs:v1-AQAB/x-overlay-layer:precomputed:\%2F\%2Fhttps:\%2F\%2Fdata-proxy.ebrains.eu\%2Fapi\%2Fv1\%2Fpublic\%2Fbuckets\%2Ftanner-test\%2FfullSharded_v1}{this URL}.}
\label{fig:siibraExplorer} 
\end{figure}

The conversion to sharded chunks instead of raw chunks, both of chunk size $\num{64} \times \num{64} \times \num{64}$ voxels, reduced the number of required files for six resolutions (\num{0.65} to \SI{20.8}{\micro\meter}) substantially from \num{20312064} to \num{4959}.
Intensity values were stored as \num{16}-bit integers. These can be converted to linear attenuation coefficients by mapping the grayscale intensity range $[\num{0}, \num{65535}]$ to $[\SI{-0.01}{\per\milli\meter}, \SI{0.10}{\per\milli\meter}]$. Note that negative attenuation values are the result of edge enhancement that was not completely removed and the air attenuation distribution being centered at zero.
The gzip-compressed pre-computed data in \texttt{Neuroglancer} format required a total of \SI{5.4}{\tera\byte} disk storage space.
The complete multi-resolution data has been uploaded to a publicly accessible \textsc{Ebrains} repository~\cite{humbel2023ebrainsFullMouseBrainData} and can be viewed via \texttt{siibra-\nolinebreak[0]explorer} using
\href{https://atlases.ebrains.eu/viewer/\#/a:juelich:iav:atlas:v1.0.0:2/t:minds:core:referencespace:v1.0.0:265d32a0-3d84-40a5-926f-bf89f68212b9/p:minds:core:parcellationatlas:v1.0.0:05655b58-3b6f-49db-b285-64b5a0276f83/@:0.0.0.-W000.._Ycgy.2-QZxH._cBFA.ys8lC..mr0..7Zz9.6X54.J0k..1cY/x-overlay-layer:precomputed:\%2F\%2Fhttps:\%2F\%2Fdata-proxy.ebrains.eu\%2Fapi\%2Fv1\%2Fpublic\%2Fbuckets\%2Ftanner-test\%2FfullSharded_v1}{this URL}
and in \texttt{Neuroglancer} using
\href{https://neuroglancer-demo.appspot.com/\#!{"dimensions":{"x":[0.0000104,"m"],"y":[0.0000104,"m"],"z":[0.0000104,"m"]},"position":[353.5,374.5,795.5],"crossSectionScale":1,"projectionScale":2048,"layers":[{"type":"image","source":"precomputed://https://data-proxy.ebrains.eu/api/v1/public/buckets/tanner-test/fullSharded_v1","tab":"rendering","shaderControls":{"normalized":{"range":[12874,25246]}},"name":"warped microCT lowres level"},{"type":"segmentation","source":"precomputed://https://data-proxy.ebrains.eu/api/v1/public/buckets/tanner-test/annotation_25_uint16","tab":"source","selectedAlpha":0.1,"notSelectedAlpha":0.5,"segments":[],"name":"Allen Atlas Segmentation"}],"selectedLayer":{"layer":"warped microCT lowres level"},"layout":"4panel"}}{this URL}.\footnote{Note, the \textsc{Ebrains} repository is under embargo until acceptance of this manuscript. Currently the URL links to view the data in \texttt{siibra-explorer} and \texttt{Neuroglancer}, stated in the text and in the caption of Figure~\ref{fig:siibraExplorer}, point to open data including only the two lowest resolutions at \num{10.4} and \SI{20.8}{\micro\meter} voxel size. These URL links will be updated to the high resolution data in the final version of the manuscript.}
Note that both viewers allow encoding the position when sharing a URL, in this case located in the caudoputamen.
Six resolution levels, in conjunction with the pre-computed \texttt{Neuroglancer} format with data chunks of $\num{64} \times \num{64} \times \num{64}$ voxels, allow fast interactive viewing via \texttt{siibra-explorer} or \texttt{Neuroglancer} by loading only the required chunks at the appropriate image resolution.
This eases navigation to microanatomical details and keeping the macroscopic overview.
Figure~\ref{fig:siibraExplorer} illustrates the visualization of the warped microtomography image in \texttt{siibra-explorer}. The correspondence between the microtomography volume and the two-photon template image of the Allen Mouse Brain CCFv3 can be observed. The microtomography image structures can be explored by navigating to predefined brain regions of the Allen Mouse Brain CCFv3 such as the parabrachial nucleus.

\section{Conclusion}
\label{sec:conclusion} 

Virtual histology based on extended-field X-ray microtomography provides isotropic micrometer resolution with fields-of-view sufficiently large to encompass more than an entire mouse brain. The dataset presented here provides rich microanatomical detail and forms a basis for future analyses. Overcoming barriers in sharing, viewing, navigating, and combining these data with existing resources will unlock collaborations with domain experts who may have limited experience in handling teravoxel-sized imaging data.
In this study, we introduced a dedicated data processing pipeline to manage terabyte-sized virtual histology datasets and integrate them with open-source interactive viewers. The reconstructed tomography data were registered and transformed into a common coordinate system, then made accessible through an online repository in a hierarchical format for exploration via web browser-based viewers. Crucially, the data's compatibility with tools like \texttt{siibra-explorer} and \texttt{Neuroglancer} allows users to apply existing labels from other datasets to rapidly identify areas of interest, overlay the microtomography volumes with images from other modalities, and share specific location coordinates via URL, facilitating collaboration with peers.

\section{Experimental Section}
\label{sec:materialsmethods}

\threesubsection{Preparation of the mouse brain}
The imaged mouse brain came from a wild-type twelve-week-old female C57BL/6JRj mouse (Janvier Labs, Le Genest-Saint-Isle, France). Tissue collection was approved by the veterinary office of the Canton of Bern (license BE98/2020). The mouse was transcardially perfused with \SI{4}{\percent} paraformaldehyde (PFA; Merck, Darmstadt, Germany) / phosphate buffered saline (PBS) pH \num{7.4} under Ketamine/Xylazine anesthesia, then the brain was dissected and immersed in \SI{4}{\percent} PFA / PBS. For measurement, the brain was dehydrated in ethanol, which has been shown to enhance tissue contrast \cite{topperwienContrastEnhancementVisualizing2019b,rodgersVirtualHistologyEntire2021a,rodgersVirtualHistologyEntire2022a,brunet_preparation_2023}. Dehydration was performed by two-hour immersion in 20 mL of \SI{50}{\percent}, \SI{70}{\percent}, \SI{80}{\percent}, \SI{90}{\percent} and \SI{100}{\percent} ethanol (Carl Roth GmbH, Karlsruhe, Germany). 

\threesubsection{Extended-field microtomography}
The mouse brain was imaged at the \textsc{Anatomix} beamline at Syn\-chro\-tron \textsc{Soleil} (Saint-Aubin, France) \cite{weitkampTomographyBeamlineANATOMIX2017,weitkampMicrotomographyANATOMIXBeamline2022}.
The undulator X-ray source was operated at a magnetic gap of \SI{5.5}{mm}. The polychromatic X-ray beam was spectrally conditioned by a succession of transmission filters with cumulative thicknesses of \SI{600}{\micro\meter} diamond, \SI{26}{\micro\meter} Au and \SI{100}{\micro\meter} Cu, resulting in an effective mean photon energy of around \SI{38}{\kilo\electronvolt}.
Projections were recorded with effective pixel size of \SI{0.65}{\micro\meter} on a sCMOS camera (Hamamatsu Orca Flash 4.0 V2, $\num{2048} \times \num{2048}$ pixels, \SI{6.5}{\micro\meter} physical pixel size) coupled to a \num{20}-\si{\micro\meter}-thick LuAG scintillator via microscope optics with \num{10}$\times$ magnification (NA \num{0.28}, Mitutoyo Corporation, Kanagawa, Japan) \cite{desjardinsDesignIndirectXRay2018a}. An exposure time of \SI{100}{\milli\second} was selected to half fill the detector's dynamic range, thus avoiding saturated pixels and reducing the severity of ring artifacts. In order to exploit propagation-based phase contrast, the detector was placed \SI{50}{\milli\meter} downstream of the sample. This was around the critical near-field distance to ensure that propagation-based blurring does not exceed two pixels \cite{weitkampANKAphaseSoftwareSingledistance2011a}.

The detector's FOV was extended by a factor of eight both perpendicular to and along the axis of rotation, see Figure~\ref{fig:acquisition_rendering} (a-c). At each height, four \ang{360}-acquisitions were recorded with \num{9000} projections and offset rotation axis to allow for about \num{200} pixels of overlap. This scheme was repeated for eight height steps with about \num{200} pixels of overlap. Each individual acquisition took around \num{15} minutes using continuous rotation mode, resulting in a total scan time of about eight hours. The tilt of the rotation stage must be aligned to a precision better than \SI{70}{\micro\radian}, as a reconstructed slice is only \SI{0.65}{\micro\meter} thick but \SI{10}{\milli\meter} wide.

\threesubsection{Reconstruction of extended field-of-view tomography data}
The processing of the extended-field microtomography was divided into
three steps, (i) determination of the rotation axis and projections
stitching positions, (ii) blending of extended projections, phase
retrieval, and ring artifact correction, (iii) tomographic
reconstruction.  These steps were implemented to be run as batch jobs,
either locally or on scientific computing infrastructure such as
sciCORE~\cite{scicore} at the University of Basel, and were controlled
by a parameter file.  For the stitching of $8 \times 8$ images to one
full projection per angle, the translational offsets between neighboring
tiles were determined globally and then applied to each angle.  The
translations between scans derived from the recorded motor positions
were taken as an initial estimate.  The offset positions were further
refined for projection sets at ten equidistant angles.  For pairs of
adjacent projections, the offset maximizing the normalized correlation
coefficient in the overlapping region of the two frames was determined.
The values obtained for ten angles were consolidated using a weighted
mean with respect to correlation peak prominence.  With the thus
determined offsets, extended projections were assembled. For overlapping
regions the image intensities of adjacent tiles were combined with
linear blending, where the original intensities $I_1$, $I_2$ are mapped
to a linear combination $I_\text{blend} = \alpha I_1 + (1-\alpha) I_2$,
with $\alpha \in [0, 1]$ denoting the position in the overlap region.
Propagation-based phase retrieval was applied with $\delta/\beta = 140$
\cite{paganinSimultaneousPhaseAmplitude2002b}. This parameter was chosen
based on visual inspection of trial reconstructions, striking a balance
between contrast and spatial resolution
\cite{rodgersOptimizingContrastSpatial2020b}.  Note that the filter was
modified to remove multiplication by the factor $1/\mu$ (see
Equation~(3) in Weitkamp et al.~\cite{weitkampANKAphaseSoftwareSingledistance2011a}). This enables
the interpretation of reconstructed values as linear attenuation
coefficient.  For ring artifact correction, the mean of
flat-field-corrected projections was taken over all rotation angles. A
high pass filter was applied to isolate inhomogeneities leading to ring
artifacts. This filtered mean projection was subtracted from all
projections.  The flat-field corrected, phase-retrieved, and blended
projections were stored in Tagged Image File Format (TIFF) with tiling
in 32-line strips for quick block-wise reading
\cite{TIFF60Specification1992}. Overall, \num{4495} projections were
produced, each measuring $\num{14982}\times\num{14982}$ pixels.
Tomographic reconstruction was performed using the \emph{gridrec}
algorithm
\cite{dowdDevelopmentsSynchrotronXray1999,
riversTomoReconHighspeedTomography2012}
implemented in \texttt{tomopy} (version 1.4.2)
\cite{gursoyTomoPyFrameworkAnalysis2014a} on blocks of 32 sinograms.
The reconstructed slices were rescaled to 32-bit signed integer range
representing attenuation values spanning $[\SI{-0.01}{\per\milli\meter},
\SI{0.10}{\per\milli\meter}]$ and stored in TIFF format.

The spatial resolution was estimated using the Fourier shell correlation (FSC)
\cite{harauzExactFiltersGeneral1986}.
A $154 \times 1020 \times 1020$ voxel volume located in the overlap of height steps three and four was reconstructed from two independent sets of projections. The two resulting volumes were multiplied with a Hamming window \cite{blackmanMeasurementPowerSpectra1958} to remove discontinuities at the borders \cite{nieuwenhuizenMeasuringImageResolution2013,shakerPhasecontrastXrayTomography2021}. The FSC curve was smoothed by applying a third order Savitzky-Golay filter \cite{savitzkySmoothingDifferentiationData1964} with window length 50. The spatial resolution, defined as the width of the smallest resolvable line pair, was determined as the inverse of the crossover frequency of the smoothed FSC with the $3\sigma$-curve \cite{vanheelFourierShellCorrelation2005}. This resulted in a spatial resolution of \SI{1.7}{\micro\meter} (full modulation period).

Volume renderings from the reconstructed data were made with \texttt{VGStudio MAX 2.1} (Volume Graphics, Heidelberg, Germany) and are shown in Figure~\ref{fig:acquisition_rendering}.

\threesubsection{Registration to Allen Mouse Brain Common Coordinate Framework}
Three-dimensional registration of large images requires substantial computational resources. For example, full mouse brain images had to be downsampled to \num{9.3}-\si{\micro\meter}-wide voxels and the number of transformation parameters had to be limited to one displacement vector per $\num{12}^3$ voxels to fit into the \SI{144}{\giga\byte} memory of the used workstation~\cite{rodgersNonrigidRegistrationDetermine2021b,rodgersVirtualHistologyEntire2022a}.

To address this challenge, a distributed hierarchical method involving sub-volume registration was developed for application at full resolution. The process is  illustrated in Figure~\ref{fig:registration}.
Initially, the volumes were downsampled to a size that allows them and the associated transformation files to be stored in memory during the image registration process. 
Subsequently, image registration was conducted for these low-resolution volumes. The resulting spatial transformation was employed to define corresponding local regions (Figure~\ref{fig:registration}, left, middle row). Axes-aligned regions of interest in the full-resolution volumes can then be processed independently. This involves transforming them in accordance with the low-resolution registration outcomes, as illustrated in Figure~\ref{fig:registration} (bottom row), and/or registering them in instances where the degree of downsampling prevented accurate alignment of fine anatomical structures.

This pipeline was employed for registering the microtomography volume to the template image of the Allen Mouse Brain CCFv3 \cite{wangAllenMouseBrain2020}, namely a hemisphere symmetric population average volume generated from two-photon microscopy images of \num{1675} mice brains. For image registration, we used the open-source software \texttt{elastix}\cite{kleinElastixToolboxIntensityBased2010c,shamoninFastParallelImage2014a} (version 5.0), as it offers standard registration functionalities as well as further useful features such as incorporation of landmark correspondences in the optimization function and definition of image regions to be registered.

The tomography data was downsampled by a factor of $32\times32\times32$, i.e., to isotropic voxels of \SI{20.8}{\micro\meter} size, to approximate the resolution of the atlas, which has an isotropic voxel size of \SI{25.0}{\micro\meter}.
Volumes were first manually aligned via a similarity transformation using \texttt{ITK-SNAP}\cite{yushkevich2016itk} (version 3.8.0). To tune and guide registration, two observers each manually selected \num{35} corresponding landmark pairs $\mathbf{L}_\textrm{F}$ and $\mathbf{L}_\textrm{M}$, where subscripts $\textrm{F}$ and $\textrm{M}$ denote fixed and moving image space, respectively.
These landmarks from both observers were split into two equally sized sets to perform two-fold cross-validation to determine suitable registration meta-parameters. Besides the image volumes, i.e.\ the fixed atlas template $\mathbf{V}_\textrm{F}$ and the moving microtomography $\mathbf{V}_\textrm{M}$, the usefulness of including image gradient magnitudes $|\nabla \mathbf{V}_\textrm{F}|$ and $|\nabla \mathbf{V}_\textrm{M}|$, as well as full brain segmentations $\mathbf{S}_\textrm{F}$ and $\mathbf{S}_\textrm{M}$ in a multi-image, multi-metric registration was tested. Thus, the spatial transformation $T$ was determined by minimizing the cost function $C$:
\begin{eqnarray} 
C & = & f_{-\textrm{MI}}(\mathbf{I}_\textrm{F},T(\mathbf{I}_\textrm{M}))+w_\textrm{G} f_\textrm{MSD}(|\nabla \mathbf{V}_\textrm{F}|,T(|\nabla \mathbf{V}_\textrm{M}|)) + \nonumber\\
 &  & w_\textrm{S} f_{-\textrm{KS}}(\mathbf{S}_\textrm{F},T(\mathbf{S}_\textrm{M}))+w_\textrm{L} f_\textrm{D}(T(\mathbf{L}_\textrm{F}),\mathbf{L}_\textrm{M})+w_\textrm{T} f_\textrm{BE}(T) \nonumber
\end{eqnarray}
The cost function $C$ includes terms for negative mutual information $f_{-\textrm{MI}}$, mean squared difference $f_\textrm{MSD}$, negative Kappa statistics $f_{-\textrm{KS}}$, and mean Euclidean distance $f_\textrm{D}$, as well as a bending energy penalty $f_\textrm{BE}$ for regularization. To include the brain boundaries, image dissimilarity measures $f_{-\textrm{MI}}$, $f_\textrm{MSD}$ and $f_{-\textrm{KS}}$ were determined for an extended brain region, i.e.\ within $\mathbf{S}_\textrm{F}$ and $\mathbf{S}_\textrm{M}$ dilated by a sphere of radius ten voxels. We tested the usefulness of incorporating $|\nabla \mathbf{V}|$ and $\mathbf{S}$ in the registration, since the image gradient magnitudes will guide the local alignment of edges and the segmentations will support global alignment.
The image dissimilarity measures were chosen to account for the nature of the image relationships. Firstly, negative mutual information $f_{-\textrm{MI}}$ accounts for the nonlinear statistical relationship between X-ray microtomography and the CCFv3 atlas template's two-photon microscopy intensity values. Secondly, mean squared difference $f_\textrm{MSD}$ ensures edge alignment. Thirdly, negative Kappa statistics $f_{-\textrm{KS}}$ maximizes agreement between binary segmentations.

Registration was based on manual alignment with a similarity transform, followed by automatic affine and then deformable registration using a grid of control points with a spacing of $s$ voxels interpolated with cubic B-Splines. 
Based on initial tests and the relative values of the cost functions, the registration meta-parameters were tested in the following ranges:
$w_\textrm{$|\nabla \mathbf{V}|$}$ $\in$ \{\num{0}, \num{e-8}\}, 
$w_\textrm{S}$ $\in$ \{\num{0}, \num{1}\},  
$w_\textrm{L}$ $\in$ \{\num{0}, \num{0.1}, \num{1}\}, 
$w_\textrm{T}$ $\in$ \{\num{1}, \num{10}, \num{100}, \num{1000}, \num{10000}, \num{100000}\}, 
and $s$ $\in$ \{\num{32}, \num{16}\}.
Finally, the images were registered using 70 landmarks and registration meta-parameters providing the lowest mean error for both cross-validation folds.

The full resolution tomography volume (moving volume) was warped to the atlas (fixed volume) by dividing the fixed space into $\num{12} \times \num{12} \times \num{12}$ target subregions, such that the axes-aligned extended moving image subregion (purple bounding box in bottom row of Figure~\ref{fig:registration}), the moving image transformed subregion (blue box in bottom row of Figure~\ref{fig:registration}), and the transformation parameters all fit in memory for an isotropic target voxel resolution of \SI{0.65}{\micro\meter}.

\threesubsection{Visualization and sharing of TB-sized data}
For dissemination and exploration of the registered data, \texttt{siibra-explorer}  and \texttt{Neuro\-glancer} were used. \texttt{Neuroglancer} is an open-source browser-based interactive visualization platform that allows for datasets up to petabyte size~\cite{neuroglancer}. The warped image regions were transformed to gzip-compressed chunks of $\num{64} \times \num{64} \times \num{64}$ voxels in sharded pre-computed \texttt{Neuroglancer} format at six resolution levels, i.e.\ \num{0.65}, \num{1.3}, \num{2.6}, \num{5.2}, \num{10.4} and \SI{20.8}{\micro\meter} for efficient visualization. The \href{https://github.com/google/neuroglancer/blob/master/src/datasource/precomputed/sharded.md}{sharded format} combines all chunks into a fixed number of larger shard files to avoid the performance penalties incurred by many small files. The sharded chunks were produced using the open-source software \texttt{Igneous} (version 4.19.2)~\cite{silversmith2022igneous}. The pre-computed data were uploaded to an \textsc{Ebrains} repository, where they can be publicly accessed with \texttt{siibra-explorer} or the \texttt{Neuroglancer} viewer using the corresponding URL. In \texttt{siibra-explorer}, users can navigate to more than \num{1300} predefined Allen Mouse Brain CCFv3 regions by their anatomical names using the underlying hierarchical ontology~\cite{wangAllenMouseBrain2020}. The data can also be accessed and processed by suitable software including
\texttt{CloudVolume}~\cite{cloudvolume} in combination with  
\texttt{Igneous}.

\medskip
\textbf{Acknowledgements} \par
The authors thank Dr.~G.~Enzmann and Prof.~Dr.~B.~Engelhardt of the Theodor Kocher Institute, University of Bern, Switzerland, for providing the animals, performing the perfusion, and dissecting the brains.
The authors thank the curation team of \textsc{Ebrains} and X.~Gui from the Forschungszentrum J\"{u}lich GmbH, Germany, for their prompt and efficient support in curating and integrating the data to the \texttt{siibra-\allowbreak explorer}.

Access to the \textsc{Anatomix} beamline was granted via Synchrotron \textsc{Soleil} proposal 20201710.
\textsc{Anatomix} is an Equipment of Excellence (EQUIPEX) funded by the Investments for the Future program of the French National Research Agency (ANR), project NanoimagesX, grant no.~ANR-11-EQPX-0031. Calculations were performed at \href{http://scicore.unibas.ch/}{sciCORE} scientific computing center at University of Basel, Switzerland. V.K.\ acknowledges support from the Swiss National Science Foundation via NCCR Kidney.CH and project no.~182683. V.K.\ and M.G.\ acknowledge  financial support from the Swiss National Science Foundation project no.~213535. M.H., C.T., B.M. and G.R.~acknowledge financial support from the Swiss National Science Foundation project no.~185058.

\medskip\pagebreak[0]
\textbf{Conflict of Interest} \par
The authors declare no conflicts of interest.

\medskip
\textbf{Data Availability Statement} \par
The data presented in this article are publicly available via the \textsc{Ebrains} repository with \href{https://doi.org/10.25493/759K-JVU}{DOI 10.25493\slash 759K-JVU}.

The source code for software used for tomographic reconstruction is available at the Github repository \href{https://github.com/unibas-bmc/mosaicreconstruction}{mosaicreconstruction}. The source code for software used for large volume transformations is available at the Github repository \href{https://github.com/unibas-bmc/LargeVolumeTransformix}{ LargeVolumeTransformix}. The downsampled microtomography dataset, masks, landmarks, registration parameters, transformation files as well as the scripts used for formatting the transformed volume for the browser-based viewers are available at the Zenodo repository
\href{https://doi.org/10.5281/zenodo.10992464}{DOI 10.5281\slash zenodo.\allowbreak 10992464}.
These resources can also be reached via the FABRIC4 portal at
\href{https://doi.org/10.5281/zenodo.11234384}{DOI 10.5281\slash zenodo.\allowbreak 11234384}.

\medskip

\bibliographystyle{MSP}
\bibliography{MousebrainManuscript}

\end{document}